\documentclass{rspublic}

\usepackage{graphicx}

\def\CE{C$_4$E$_1$}

\begin{document}

\title[Ramped Phase Separation]{Phase Separation in Binary Fluid Mixtures with Continuously Ramped Temperature}

\author[M. E. Cates, J. Vollmer, A. Wagner, D. Vollmer] 
{M. E.  Cates$^1$, J. Vollmer$^{2}$, A. Wagner$^{1,3}$, D. Vollmer$^{1,4}$}

\affiliation{ 
  $^1$ School of Physics, University of Edinburgh, King's Buildings, Edinburgh EH9 3JZ, UK\\
  $^2$ Max-Planck Institute for Polymer Research, Ackermannweg 10, 55128 Mainz, Germany\\
  $^3$ Department of Physics, North Dakota State University, Fargo, ND 58105, USA \\
  $^4$ Institute for Physical Chemistry, University of Mainz, Welder-Weg 11, 55099 Mainz, Germany}

\label{firstpage}

\maketitle

\begin{abstract}{demixing, gravity, temperature ramp, binary fluids}
  We consider the demixing of a binary fluid mixture, under gravity,
  which is steadily driven into a two phase region by slowly ramping
  the temperature. We assume, as a first approximation, that the
  system remains spatially isothermal, and examine the interplay of
  two competing nonlinearities. One of these arises because the
  supersaturation is greatest far from the meniscus, creating
  inversions of the density which can lead to fluid motion; although
  isothermal, this is somewhat like the B\'{e}nard problem (a single-phase
  fluid heated from below). The other is the intrinsic diffusive
  instability which results either in nucleation or in spinodal
  decomposition at large supersaturations.  Experimental results on a
  simple binary mixture show interesting oscillations in heat capacity
  and optical properties for a wide range of ramp parameters. We argue
  that these oscillations arise under conditions where both
  nonlinearities are important.
\end{abstract}

\section{Introduction}

If a binary fluid, made of two species that are miscible at high
temperature, is suddenly quenched into the two-phase region, it starts
to demix. Depending on composition and quench depth, the mechanism for
demixing and subsequent domain growth is either spinodal decomposition
(amplification of small compositional fluctuations, at first by
diffusion and then by fluid motion driven by interfacial tension) or
nucleation and diffusive growth of small droplets of one phase in the
other (Bray 1994, 2000; Onuki 2002). Unless the two fluids have
exactly matched densities, gravity eventually takes over, once
the domain (or droplet) size becomes comparable to a suitably defined
capillary length (Onuki 2002). The details of this gravitational stage
are not fully understood and involve interesting new physics such as
`lane formation' (Chan \& Goldberg 1975, Aarts \textit{et al.} 2002).
Nonetheless, it is observed that, once gravity does intervene, fluids
separate relatively rapidly leading finally to a flat horizontal
meniscus between phases.

In the natural world, and even in most laboratory settings, rapid
temperature changes are a relatively rare occurrence. The opposite
case of a very slow temperature ramp, though equally idealised, is
arguably closer to most everyday instances of phase separation, and it
is certainly important to understand this limiting case. However, the
physics is more complex, because the system is now subject to
continuous driving, as opposed to being displaced instantaneously from
equilibrium, and allowed to relax back towards it.

If the ramp rate $dT/dt$ is small enough, one can hope that the time
taken for heat to diffuse across the sample is small compared to other
time scales of interest\footnote{ 
  This time $\tau_h$ obeys $\Lambda_y^2 \sim \kappa_T \tau_h$ where
  $\Lambda_y$ is the smallest dimension of the sample cell (its thickness) and $\kappa_T$ the thermal
  diffusivity.}.
If so, this allows us to treat the system as isothermal at any
instant; for simplicity we shall do this here, although in the
experiments that motivate this work, the spatial gradients of
temperature may not in fact be negligible. Neglecting all such
gradients reduces the problem from a double-diffusive one (Brandt \&
Fernando 1995), where diffusion of heat and composition are both
important, into one involving compositional diffusion only.

Despite this the problem is still `doubly nonlinear'. The first
nonlinearity is standard, and arises from the coupling between density
differences (caused by composition deviations) to gravity. But the
fact that the system can show phase separation requires that, even
without this coupling, compositional diffusion is already a highly
nonlinear process: a linearized diffusion equation obviously cannot
yield spinodal decomposition and/or nucleation of droplets. The
interplay of these two nonlinearities, for the case of a slow
isothermal temperature ramp in a binary fluid system undergoing phase
separation, is addressed in this paper.

\section{Experimental motivation}

Vollmer \textit{et al.} (2002) reported differential calorimetric
studies of the phase separation of a binary fluid mixture comprising
water and 2 butoxy ethanol, also called \CE, at low ramp rates (a few
Kelvins per hour). The system has a lower consolute point and
therefore demixes when heated up.
The experiments revealed a quite unexpected effect: after an
incubation period during which the phase boundary is first crossed and
a meniscus forms, the observed heat capacity passes through a series
of oscillations (typically about six in number) before eventually
decaying to a smooth curve (Figure 1, upper left). The period and
overall duration of the oscillations depends on heating rate; it also
depends on sample geometry, and on the composition of the initial
state.

An experiment in which heating was stopped for several hours midway
through the oscillations (Figure 1, upper right) showed that these
resumed when heating was recommenced, with only a small time delay;
their period was not affected. Moreover, the eventual decay of the
oscillations occurred exactly as though no interruption had occurred.
These observations establish that there is little inertia involved in
the process (since the interruption brings all fluid motion to a
halt), and also suggest that the final decay of the oscillations is
intrinsic, rather than stemming (for example) from a gradual loss of
coherence in different parts of the system. (The recommencement of
heating after a long break would presumably restore any required
coherence.)

\begin{figure}
\[ 
\includegraphics[width=0.45\textwidth]{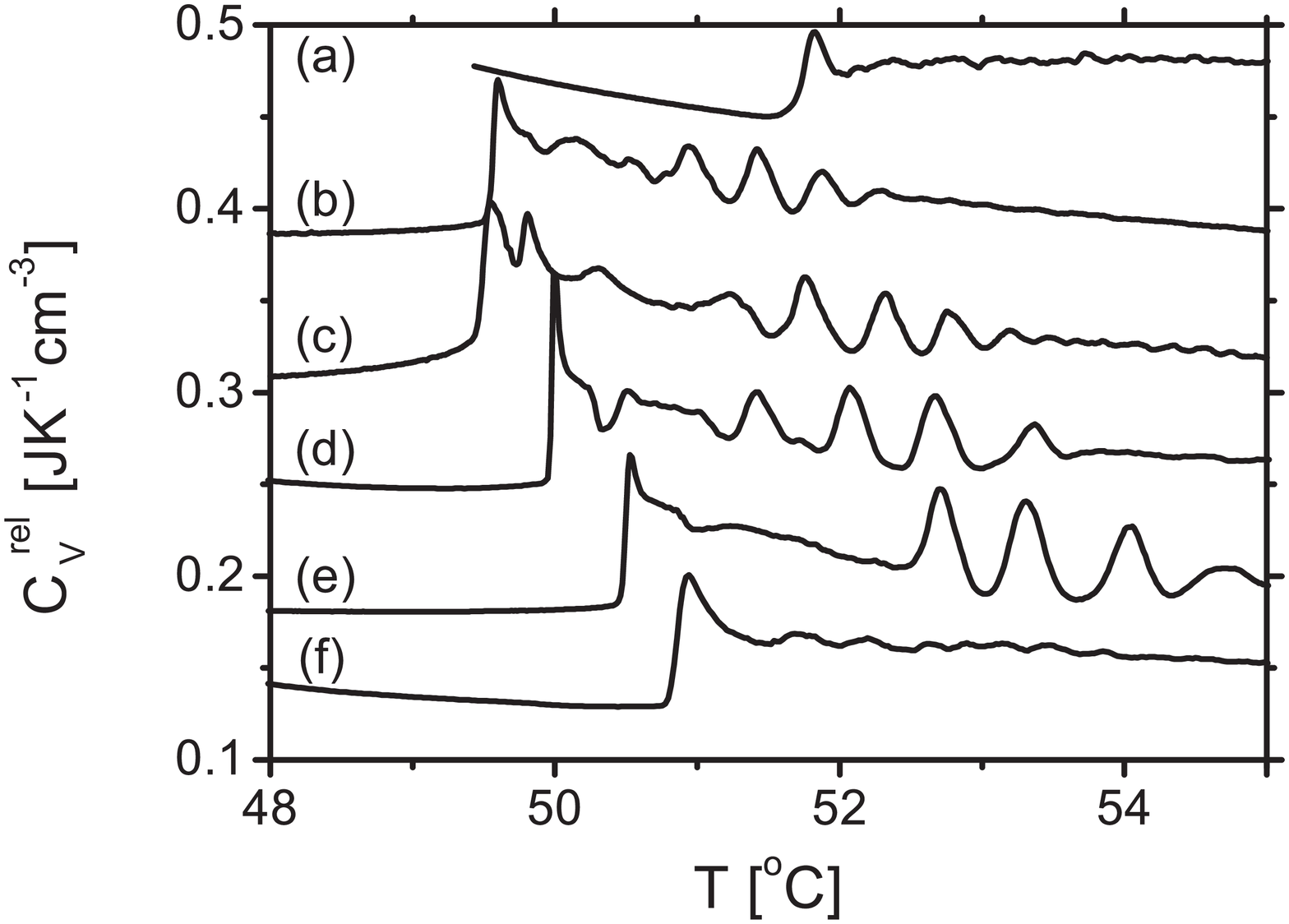}
\rule{0.1\textwidth}{0mm}
\includegraphics[width=0.45\textwidth]{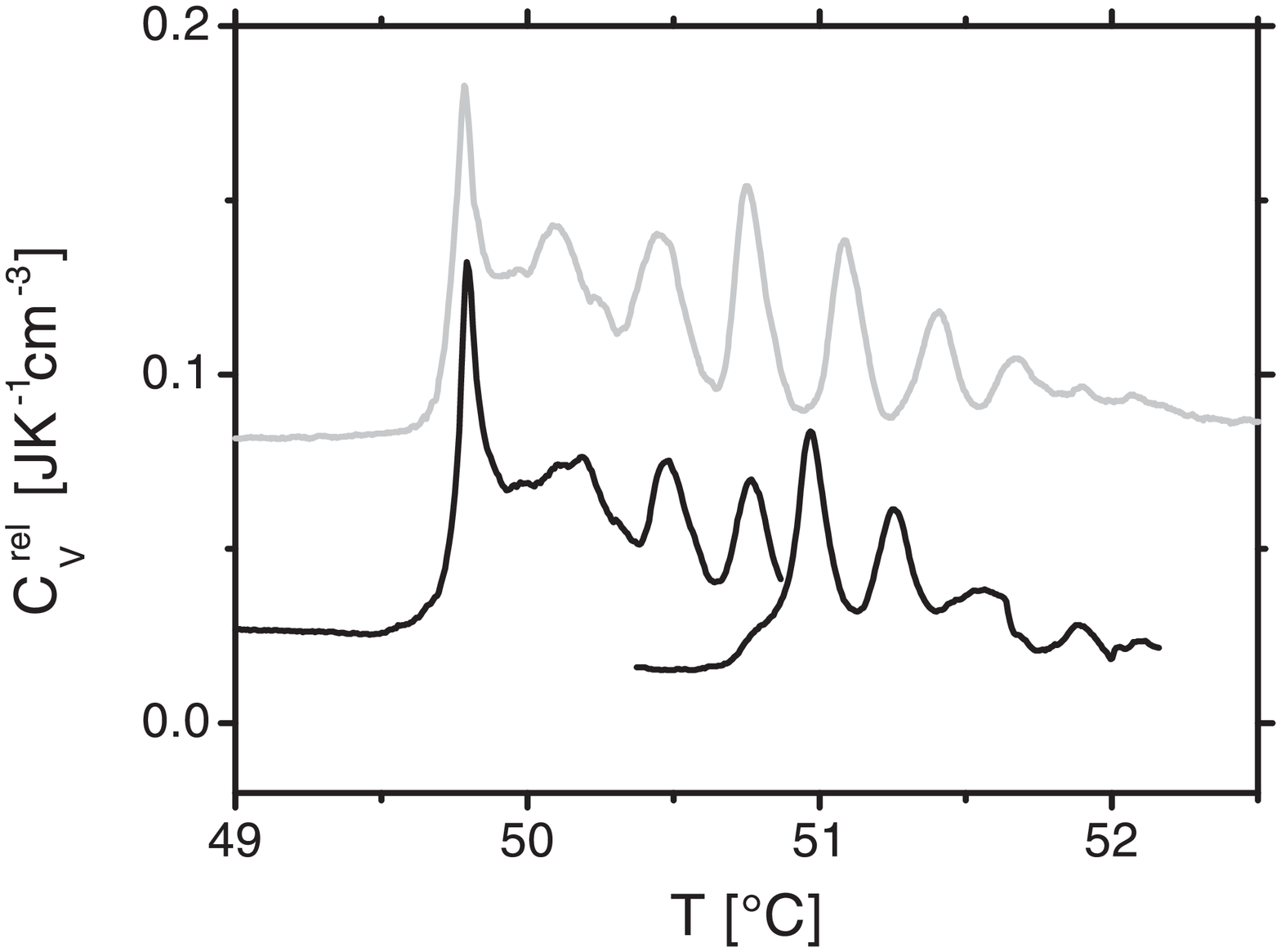} 
\]
\vspace{2mm}
\[ \rule{10mm}{0mm}
\includegraphics[width=0.8\textwidth]{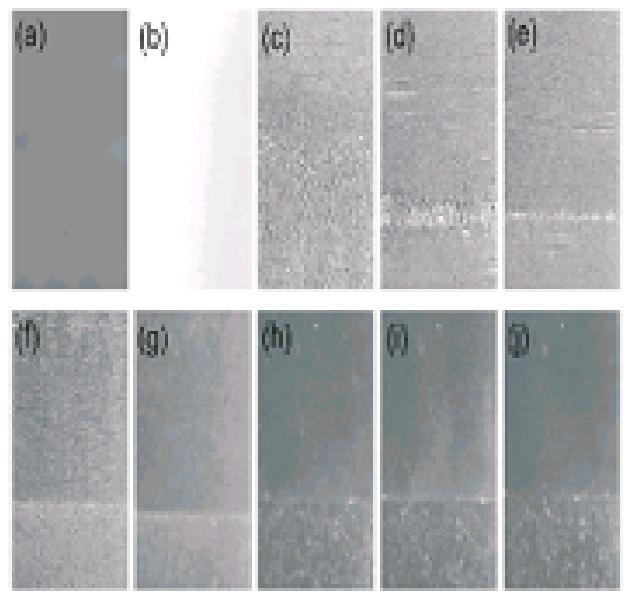} \]
\label{figure1}
\caption{
  Heat capacity measurements and videomicroscopy for mixtures of \CE\
  and water.
  \textbf{ (top left)} Apparent heat capacity of mixtures of  various initial compositions
  [(a) 0.18; %
   (b) 0.25; %
   (c) 0.35; %
   (d) 0.39; %
   (e) 0.43; %
   (f) 0.44 \% volume fraction\CE ] %
   that undergo a temperature ramp of 2~K/hour.
   \textbf{ (top right)} Comparison of a scan (0.4 \CE by volume, 0.8~K/hour ramp rate) with one
   where the heating was stopped after the
   first three oscillations and resumed at a later time (black).
   \textbf{ (bottom)} Videomicroscopy sequence for a heating rate of 10~K/hour
   and an initial composition of 32 \% \CE\ by volume. The temperatures
   ($^\circ$C) are (a) 48.8, (b) 50.5, (c) 51.1, (d) 51.2, (e) 51.3,
   (f) 52.4, (g) 53.4, (h) 57.9, (i) 58.8, (j) 59.5. The first oscillation in
   the sequence is visible in the final three frames. In frame (b) the sample is white.
   (The upper left and bottom figures are reproduced from Vollmer
   \textit{et al.} (2002).)}
\end{figure}

By observing the upper phase present in these samples using contrast
enhanced videomicroscopy (Figure 1, bottom) the oscillations were
associated with the following cycle. One observes an onset of high
turbidity as large numbers of small droplets are suddenly created;
these coarsen, and are soon cleared from the system by gravity. Bulk fluid motion is visible during this sedimentation stage;
the system is then left in a quiet, transparent state with few
droplets present. Soon after, the process repeats.
In the quiet interludes between episodes of droplet formation, it is
observed that fluid motion
continues.

\section{Theoretical Considerations}

\subsection{Equations of motion}

We introduce a minimal model which is not strictly realizable in
experiment (although it can more nearly be achieved in computer simulation; see Vollmer \textit{et al.}, unpublished). This model addresses an
isothermal, thermodynamically symmetric binary fluid, whose kinematic
viscosity $\nu$ (\textit{i.e.}, the diffusivity of momentum) is independent of composition and temperature, and
whose mass density $\rho$ depends on the composition variable $\phi$
through a temperature independent parameter 
$\alpha = \rho^{-1} \rd\rho/\rd\phi$.  
The equilibrium phase diagram is symmetric about $\phi = \phi_c = 0$,
with the compositions of the two coexisting phases (at the binodal) obeying $\phi = \pm \phi_0(T)$. The fluid is taken 
to be incompressible ($\nabla.{\bf u} = 0$). 

The Navier Stokes equation for the fluid velocity ${\bf u}$ may then
be written
\begin{equation}
\frac{\partial {\bf u}}{\partial t}
+ {\bf u}.\nabla {\bf u} 
+ \alpha {\bf u} \left(
  \frac{\partial \hat\phi}{\partial t} + {\bf u}.\nabla \hat\phi
  \right)
= \nu\nabla^2{\bf u} 
+\alpha g\phi_0 \hat\phi \nabla z 
- \nabla (p/\rho+gz)
\label{eone}
\end{equation}
Here $g$ denotes gravity, acting towards $-z$; $p$ is the fluid
pressure; and $\hat \phi = (\phi_0-\phi)/\phi_0$ is a relative
supersaturation that measures the deviation of the local composition
from the equilibrium value $\phi_0$. This holds in regions of positive
$\phi$; otherwise the sign of $\hat\phi$ is reversed. (Note that a
state of zero $\hat\phi$ can describe two phases at equilibrium, with
a sharp meniscus in between.)  In what follows, we shall assume that
the effects of composition on fluid density (proportional to $\alpha$)
mainly enter through the term in $g$ on the right hand side: this term
represents a buoyancy effect in which local composition affects
the body force acting directly on the fluid. The other
place that $\alpha$ enters, on the left of the equation, represents
the fact that changes in mass density caused by varying composition
alter the acceleration of a fluid element subject to any particular
set of forces. We shall neglect this below.

The Navier Stokes equation is coupled to the nonlinear
advection-diffusion equation (\textit{cf.} Appendix A for a derivation) 
\begin{equation}
\frac{\partial \hat\phi}{\partial t} 
+ {\bf u}.\nabla \hat\phi 
= \nabla \left( \bar{D}\,f(\hat\phi) \nabla \hat\phi \right)
+ (1-\hat\phi)\xi
\label{etwo}
\end{equation}
The comoving derivative on the left accounts for `advection': the
process whereby composition is transported by bulk fluid flow. The
first term on the right represents diffusive currents, and the second
is a source term for the supersaturation $\hat\phi$. 
In contrast to the linear diffusion equations appearing in other
settings, this equation has two remarkable features:
\\
\textbf{(i) } The variable $\xi(t)$ denotes $\phi_0^{-1}d\phi_0/dt$
and is a measure of the ramp rate. The source term, in
which $\xi$ appears, characterizes the change of composition
$\phi_0(T)$ with time during the temperature ramp, which gives rise to
a constant \emph{increase} of the relative supersaturation $\hat\phi$
under conditions where no diffusion or advection take place. As far as
$\hat\phi$ is concerned, the effect of the temperature ramp is
therefore to create supersaturation, at a rate controlled by $\xi$,
throughout the sample.
\\
\textbf{(ii) } Equation \ref{etwo} involves an effective
\emph{nonlinear} diffusivity $\bar{D} f(\hat\phi)$, which decreases
with the relative supersaturation $\hat\phi$, becoming negative for
$\hat\phi > \hat\phi_s$ (Bray 1994). In particular $\bar{D}$ is chosen
in such a way that $f(0)=1$ and $f(\hat\phi_s)=0$ at the spinodal
$\hat\phi_s$ where any state of uniform $\hat\phi > \hat\phi_s$
becomes locally unstable.  Henceforth, we assume, for simplicity, that
$\bar{D}$ and $f(\hat\phi)$ do not depend directly on temperature, so
that the only $T$-dependence of the diffusive dynamics enters
parametrically through the definition,
$\hat\phi = 1-\phi/\phi_0(T)$,  
of the relative supersaturation.

\subsection{Dimensionless parameters}

The equations of motion can be made dimensionless by introducing
suitable scales of mass, length and time.  As a length scale we choose
the height of the sample $\Lambda\equiv\Lambda_z$ as the unit of
length, and $\Lambda^2/\bar D$ as time unit. A mass unit is
$\rho\Lambda^3$ but interestingly, none is really needed.  The fluid
density in equation \ref{eone} factors through all terms in the
equation, except for the one involving the pressure,
$\nabla(p/\rho + gz)$,   
whose sole purpose is to maintain incompressibility.  We will write
this term as a dimensionless gradient $\nabla \tilde p$, but could
equally well eliminate it by projecting the Navier Stokes equation
onto incompressible flows in a standard fashion (see e.g. Onuki 2002).

With these units, allowing for the various assumptions outlined above,
and furthermore imposing a constant `ramp rate' $\xi$, we find that
the equations of motion take the following form:
\begin{eqnarray}
\frac{\partial {\bf u}}{\partial t} + {\bf u}.\nabla {\bf u} 
&=& N_1\left[\nabla^2{\bf u} + (N_0/N_2)\hat\phi \nabla z - \nabla \tilde p\right]
\label{ethree}
\\
\frac{\partial \hat\phi}{\partial t} + {\bf u}.\nabla \hat\phi 
&=& \nabla \left( f(\hat\phi) \nabla \hat\phi \right) 
+ (1-\hat\phi)N_2
\label{efour}
\end{eqnarray}
with three dimensionless parameter groups:
\begin{equation}
N_1 = \nu/\bar D \;\;\; ; \;\;\; N_2 = \xi \Lambda^2/ \bar D
\;\;\;;\;\;\;
\frac{N_0}{N_2} = \alpha g \phi_0 \Lambda^3/ \bar D\nu  \label{efive}
\end{equation}
Of these, $N_1$ is a material parameter of the binary fluid (the ratio
of momentum to particle diffusivities). $N_2$ can be thought of as a
dimensionless ramp rate (although a different one, which does not
depend on $\Lambda$, will be introduced later on). The group we have
denoted $N_0/N_2$ is, in the same sense, a dimensionless gravity
parameter. But we will find below that the product of this with $N_2$
plays a special role in the theory; that product deserves its own
name, and we call it $N_0 = \alpha g \phi_0\xi\Lambda^5/\bar D^2\nu$.
Although these three groups can be combined in numerous ways to create
new dimensionless numbers, they are sufficient to fully describe
the parameter space in our idealized problem, for a given sample shape
and given $f(\hat\phi)$.

The statement that they are sufficient assumes that no important
physics has been left out of the equations of motion \ref{ethree} and
\ref{efour}.  Perhaps the strongest candidates for missing physics are
(a) thermal gradients, which we have neglected from the outset; and
(b) interfacial tension, which arise locally once the nonlinear regime
of droplet formation is encountered, and are present at all times at
the meniscus between phases. (Appendix A describes how a tension could
be included.)

One dangerous-looking combination of these neglected effects is when
interfacial tension and thermal gradients combine to create Marangoni
stresses (that is, spatial gradients of the tension); these are known
to be implicated in several interfacial instabilities (Sternling \&
Scriven 1959, Davis 1987). While it would be imprudent to rule out
an important role for Marangoni stresses and other thermal-gradient
related effects, we do think it worth neglecting these in the first
instance.

\subsection{Relation to a B\'{e}nard problem: Advective instability}

Consider first the limiting case where $f(\hat\phi) = 1$, or
equivalently, where $\hat\phi$ remains infinitesimal. Once the
meniscus has formed, we can envisage a static diffusive state in which
a steady current of supersaturation flows from the upper phase towards
the lower and vice versa. (Note that, in such a state, $\hat\phi$
vanishes on the meniscus itself.) This current arises from the source
term $N_2$ in equation \ref{efour}, which creates supersaturation
uniformly through space. However, were such a current to come from a
localized source that maintained constant supersaturation $\hat\phi^*$
at the top of the sample, equations \ref{ethree} and \ref{efour} would
be isomorphic to those for the standard (Rayleigh-)B\'{e}nard problem (see e.g.,
Faber 1995). The latter concerns a single phase fluid heated from
below; this creates an inverted density gradient (with the denser
fluid on top) so that above a certain heating rate the system becomes
unstable. The role of the local temperature field, in the B\'{e}nard
problem, is played in ours by the supersaturation; that of the
temperature difference between the plates, by $\hat\phi^*$. The
correspondence applies whatever the sign of the expansion coefficient
$\alpha$: the less dense phase is always on top, so that
supersaturation in this phase, which is largest far from the meniscus,
always leads to a density inversion (and the same happens,
symmetrically, in the lower phase too).

The presence of a distributed source term, rather than driving at the
boundary, gives an extra complication to this analogy due to the
nontrivial form of the quiescent profile. This strongly
alters the details of any stability analysis, but not the basic ideas.
First, note that because the B\'{e}nard problem is linear in the diffusive
sector, it is governed by only two dimensionless groups, not three.
This is the case in our problem also, so long as $\hat\phi$
(controlled by $N_2$) remains small enough. The two parameters
normally chosen for the B\'{e}nard problem are the Prandtl number, which
is the direct analogue of our $N_1$, and the Rayleigh number, which in
our language is Ra $ = (N_0/N_2)\hat \phi^*$. The latter expression
pertains to the `standard' case of a constant supersaturation $\hat
\phi^*$ at the upper boundary. When the source is distributed
uniformly, this characteristic supersaturation scale instead depends
on both the ramp rate and the system size, as $\hat\phi^* \simeq
\xi\Lambda^2/\bar D = N_2$. This follows from the linearized diffusion
equation: $\hat\phi^*$ is the supersaturation that builds up at a
distance of order $\Lambda$ from the meniscus within the time it takes
for diffusive relaxation over that distance.  Combining these results,
we may therefore identify $N_0$ itself as the direct analogue of a
Rayleigh number for our variant of the B\'{e}nard problem.

Much is known about the B\'{e}nard problem, and using it we may now guess,
with reasonable confidence, what will happen as the ramp rate $\xi$ is
increased. The static diffusive profile will be stable until $N_0$
exceeds a critical value $N_0^c$ (for the standard B\'{e}nard problem,
this is about 1700); thereafter, circulating fluid rolls will arise.
This transition is represented in the phase diagram Figure 2 (left) as
the transition from the region marked S to A.
As $N_0$ increases further, these rolls will undergo a series of
bifurcations through various steady and perhaps unsteady (including
oscillatory) states, with the details of this process controlled by
$N_1$. Eventually at some much larger $N_0$ (also dependent on $N_1$),
chaotic motion will set in, leading finally to fluid turbulence.

Note in passing that, although the rolls in B\'{e}nard are normally called
`convection rolls', they are properly `advection rolls', in the
following sense. The onset of these rolls is controlled by the
advective nonlinearity (transport of supersaturation by fluid, which
is the ${\bf u}.\nabla \hat \phi$ term on the left side of equation
\ref{efour}) not the convective one (transport of momentum by fluid,
which is the ${\bf u}.\nabla {\bf u} $ in equation \ref{ethree}). For
typical fluid parameters ($N_1 \simeq 1$ or more) one must move rather
far along the sequence of transitions beyond $N_0^c$ before the
Reynolds number becomes large enough that convective, rather than
advective, nonlinearity becomes important.

\begin{figure}
\[ \includegraphics[width=0.45\textwidth]{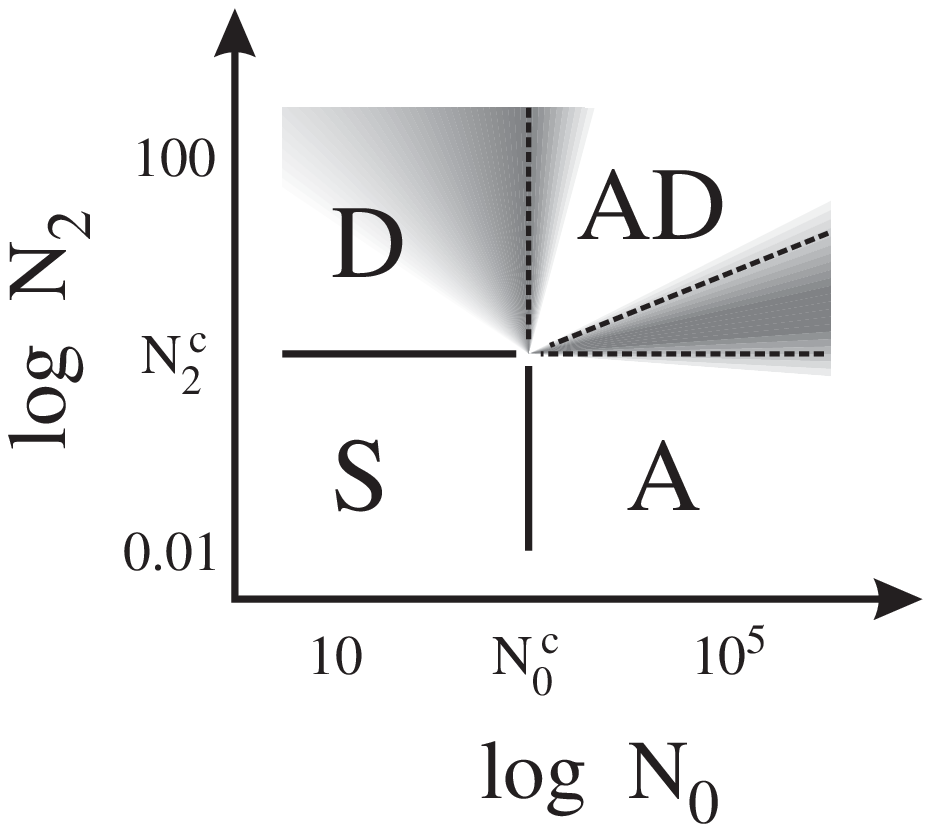}
   \rule{0.1\textwidth}{0mm} 
   \includegraphics[width=0.45\textwidth]{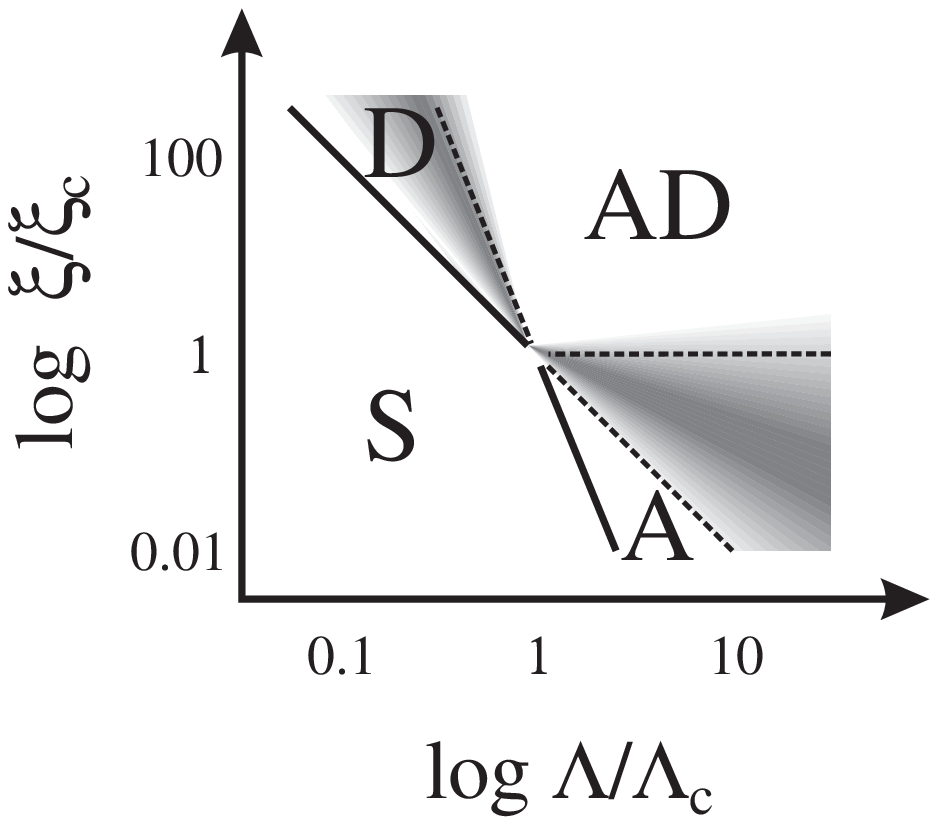} \]
\label{figure2}
\caption{ 
  Suggested schematic phase diagram for an isothermally ramped binary
  fluid. The left panel gives the phase diagram in terms of the
  dimensionless control parameters $N_0$ and $N_2$, while the right is
  given in terms of $\Lambda$ and $\xi$; both in log/log plot. 
  The regions with qualitatively different behaviour are marked as S =
  stable, A = advectively unstable, D = diffusively unstable (perhaps with steady sedimentation), AD =
  advectively and diffusively unstable. Crossover lines from D to AD and from A to AD probably lie within
  the shaded regions shown. (In practice, of course, the boundary of the S region will not really have a sharp corner but will be a smooth curve, with the straight lines shown here as asymptotes. Likewise the `crossover lines' need not meet the phase boundary precisely at this corner.)}
\end{figure}

\subsection{Onset of droplets: Diffusive instability}

All of the preceding subsection refers to the case where the
characteristic supersaturation $\hat\phi^*$ remains small enough that
the diffusion is linear.  Let us now consider the opposite case where
$N_0$ remains small, but $\hat\phi$ does not.

For small $N_0$ we expect no onset of advection rolls before the
diffusive nonlinearity kicks in. Within a linearized framework, we
found $\hat\phi^* \simeq N_2$, so the onset of the diffusive
nonlinearity, if present on its own, must be governed simply by 
$N_2 \equiv \xi\Lambda^2/\bar D$.  
When this reaches some critical value $N_2 = N_2^c$ (which in general
depends on the shape of $f(\hat\phi)$), the stationary diffusion
profile becomes unstable, which is indicated in the phase diagram
(Figure 2, left) by the transition from the region S to D. 
In region D droplets are formed by either a spinodal or a nucleation
process.  The instability must occur at or before the point where the
supersaturation in the static profile reaches $\hat\phi_s$; this
yields an upper bound $N_2^c \le N_2^{c,s}$ (at which the region of
highest supersaturation is locally unstable to spinodal
decomposition).
A particular choice of $f$ yields 
$N_2^c = N_2^{c,s}\simeq 1$  
at small enough $N_0$ (J Vollmer, unpublished), which appears to be
adequate for our present purposes. In physical terms this result
states that the diffusive instability kicks in whenever the height of
the sample exceeds the length scale $(\bar D\xi)^{1/2}$ beyond which
diffusion can no longer compete with the homogeneous growth of
supersaturation characterized by $\xi$. 
 
It is not completely clear what should happen once droplet formation
begins. If gravity is strictly zero ($N_0 = 0$) the system will
coarsen without ever forming a horizontal meniscus; the presence of
the ramp may cause domain formation within existing domains,
reminiscent of patterns seen in certain binary systems undergoing
simultaneous phase separation and polymerization (Clarke \textit{et
  al.} 1995). (This process might conceivably lead to oscillatory
states; the details certainly depend on $N_1$.) In any case, once gravity is switched on, sedimentation of
the nucleated domains will occur. By gathering supersaturation into
localized droplets, nucleation is likely to result in the onset of
advective nonlinearity {\em sooner} than would be the case without it;
that is, {\em before} the line $N_0 = N_0^c$ is crossed (gray region
in sector D). In the first instance the result could merely be steady
sedimentation of the droplets; although this is, technically, an `advective
nonlinearity' (since it involves the ${\bf u}.\nabla \hat \phi$ term in
equation \ref{efour}) it is relatively benign. By relaxing supersaturation, this could cause the threshold for bulk flow to move somewhat beyond $N_0^c$; but such sedimentation, once present, is liable
to give rise to fronts (Russel \textit{et al} 1989) and 
other collective nonlinear behaviour (Chaikin
2000), distinct from those of the B\'{e}nard problem.
In summary, the onset of advective nonlinearity (beyond simple sedimentation), in a state where the
diffusive one is already active, should occur at a threshold which is likely to lie somewhere between the diffusive instability line ($N_2 =N_2^c$)
and the continuation of the advective instability line ($N_0 =
N_0^c$).
 
By the same token, starting with $N_0 > N_0^c$ and $N_2$ small, the
presence of B\'{e}nard rolls will clearly alter the criterion for onset of
droplet formation. Here one can argue that the B\'{e}nard rolls are more
efficient than pure diffusion at transporting supersaturation, so that
the onset of the diffusive instability will be delayed {\em beyond}
the line $N_2 = N_2^c$ as was calculated for small $N_0$. On the other
hand, for a given ramp rate $\xi$ (and fixed material parameters), we
can find the cell size $\Lambda^* = \Lambda(N_0^c/N_0)^{1/5}$ for
which the B\'{e}nard instability would be just incipient. It would be
surprising if, in a sample of size $\Lambda \gg \Lambda^*$, the
supersaturation far from the meniscus were actually less (for the same
ramp rate and material parameters) than the value 
$\hat\phi^*\simeq N_2(\Lambda^*/\Lambda)^2$  
pertaining to this smaller cell. This suggests that diffusive
instability should set in {\em at the latest} upon achieving the
condition $N_2(N_0^c/N_0)^{2/5} = N_2^c$, where the left side of the
equation is the value of $N_2$ in a fictitious sample of height
$\Lambda^*$. 
In summary, the onset of diffusive nonlinearity, in a state where the
advective one is already active, should occur somewhere in the shaded
region marking the transition from the A to the AD regions in the left
diagram of Figure 2, whose boundaries are the continuation of the
diffusive instability line ($N_2 = N_2^c$) and the line
$N_2/N_2^c \sim (N_0/N_0^c)^{2/5}$.

\subsection{Nonequilibrium phase diagram for experimental control parameters}

The above arguments, which were summarized in the left panel of Figure 2, predict a phase boundary between a stable diffusive
regime (S) and an advectively unstable one (A) at $N_0 = N_0^c$; and a boundary
between S and a diffusively unstable regime (D) at $N_2 = N_2^c$.  For
comparison with experiment it is helpful to represent these lines on a plot that involves two more accessible
parameters, the system size $\Lambda$ and ramp rate $\xi$, as the horizontal
and vertical axes (Figure 2, right). To this end we observe that the
combinations 
$N_0/N_2 = (\Lambda/\Lambda_0)^3$ and  
$N_2^5/N_0^2 = (\xi/\xi_0)^3$ only depend on material constants. In these expressions we have identified `natural' units of length and heating rate 
$\Lambda_0^3 = \bar{D} \nu / \alpha g \phi_0$ and  
$\xi_0^3 = \alpha^2 g^2 \phi_0^2 \bar{D}/\nu^2 =  (\bar{D} / \Lambda_0^2)^3$,  
respectively, only depend on material constants. 
Consequently, on logarithmic scales the phase diagrams can be related
by the linear transformation
\begin{equation}
\left(\begin{array}{l}
  \log\frac{\Lambda}{\Lambda_c} \\[1mm]
  \log\frac{\xi}{\xi_c} 
\end{array}\right)
= 
\left(\begin{array}{rr}
 \frac{1}{3} & -\frac{1}{3} \\[1mm]
-\frac{2}{3} &  \frac{5}{3}
\end{array}\right)
\left(\begin{array}{l}
  \log \frac{N_0}{N_0^c} \\[1mm]
  \log \frac{N_2}{N_2^c} 
\end{array}\right)
\end{equation}
where $\Lambda_c$ and $\xi_c$ are chosen so that the two
boundaries of the stable region S intersect at $(1,1)$, \textit{i.e.,}
the sharp corner visible in the right panel of Figure 2.
Consequently, the S/A and S/D lines now meet at the point
($\Lambda_c,\xi_c$), where 
$\Lambda_c/\Lambda_0 = (N_0^c/N_2^c)^{-1/3}$, and  
$\xi_c/\xi_0 = (N_2^c)^{5/3}(N_0^c)^{-2/3}$  
as defined previously.

The estimates for the boundaries between the stable domain S and the
advectively (A) and diffusively (D) unstable ones translate in this
phase diagram into straight lines with slope -5 and -2, respectively.
The limiting estimates made above for where regimes A and D should
each cross into AD (the regime in which advective and diffusive
nonlinearities are simultaneously strong) are shown by dashed lines as
previously, and the uncertainties for these regions are again
indicated by gray areas. 
The estimate of the upper bound for the appearance of nucleation in
the advective regime $N_2 \sim N_0^{2/5}$ translates into a condition
on the ramp rate $\xi$, and is independent of the sample height
$\Lambda$ (although it will, like other critical parameters, in
general depend on the {\em shape} of the sample cell, and on
$f(\hat\phi)$).

Not shown in either phase diagram are additional, increasingly vague
crossovers from A, D and AD into further regimes where the convective
nonlinearity also becomes important.

\section{Oscillation Mechanism}

In the experiments showing oscillatory demixing in \CE, that were
outlined in \S 2,
one observes fluid flow (without obvious signs of turbulence) and
droplet formation simultaneously. This means that advective and
diffusive nonlinearities are both involved, making it plausible that
these experiments lie within the AD region of the phase diagram.
Preliminary parameter estimates 
(Vollmer \textit{et al.}, unpublished)
suggest that they may lie in a part of the AD region well to the right
of, but not much above, the point $(\Lambda_c,\xi_c)$ on Figure 2
(right).  However, in the experiment it is $dT/dt$ that is held
constant, not $\xi$, which can therefore drift slowly during the ramp,
as can $\alpha$, $\nu$, etc. These drifts may cause the experimental
parameters to move gradually towards more stable values, causing the
oscillations finally to cease.

These remarks do not explain why the oscillations are
there in the first place. Elsewhere 
we propose that in this part of the AD region, advection rolls form
soon after creation of the meniscus, while the supersaturation
$\hat\phi$ is still small. These rolls do not advect enough flux to
maintain $\hat\phi$ below the threshold for droplet nucleation. Hence
droplets appear; they then grow for a time, but before long,
sedimentation sets in. The falling droplets soak up the remaining
supersaturation, restoring the system to a state of small $\hat\phi$,
so that the cycle is ready to begin again. In Vollmer \textit{et al.}
(unpublished) we find this scenario consistent with preliminary
computer simulation data, and broadly in line with experimental data
on various systems.

This explanation has something in common with a proposed mechanism for
rhythmic deposition of precipitate in a convecting magma chamber
cooled from above (Sparks \textit{et al.} 1993), a problem that seems,
however, to involve a significantly larger parameter space than the model considered here.

\section{Conclusion}

Even within the idealized model presented here, the interplay of
gravity-driven advection and nonlinear diffusion presents an
interesting theoretical challenge for the description of binary fluid
demixing in the presence of a temperature ramp. The experimentally
observed oscillations (Vollmer \textit{et al.} 2002) appear
in a regime where both nonlinearities are simultaneously
large; this makes analysis difficult. Detailed computer simulations,
now planned, may be a better way of finding out whether the physical
ingredients retained in the idealized model (equations \ref{ethree}
and \ref{efour}) are sufficient to explain the observed oscillations,
or whether factors that we have omitted, such as Marangoni stresses
arising from thermal gradients, play an essential role in this
problem.

\begin{acknowledgements} J.V. acknowledges financial support from the Schloessmann Foundation of the Max Planck Society, and thanks Howard Stone and Michael Brenner for useful discussions. D. Vollmer thanks the EC for a Marie Curie Fellowship.
\end{acknowledgements}

\appendix{Advection-diffusion equation for the supersaturation $\hat\phi$} 

The evolution of the composition $\phi$ of the mixture is governed by
the advection-diffusion equation 
\begin{equation}
\frac{\partial \phi}{\partial t} 
+ {\bf u}.\nabla \phi = - \nabla J_\phi
\label{diffusion}
\end{equation}
where the comoving derivative on the left accounts for the advection
of composition by the bulk fluid flow; and the divergence of the
diffusive current $J_\phi$ on the right hand side accounts for the
diffusive decay of composition gradients. 
The current $J_\phi = -M\nabla\mu$, with $M$ a mobility, is driven by the
gradient of chemical potential $\mu$, which --- in local equilibrium ---
obeys $\mu = \delta F/\delta\phi$, where a suitable choice for the free
energy functional $F[\phi]$ is
\begin{equation}
F[\phi] = \int \rd V \left[ F_0(\phi) + \kappa(\nabla\phi)^2\right]
\label{esix}
\end{equation}
This involves the free energy density $F_0(\phi)$ describing homogeneous
phases in equilibrium, and an energetic penalty $\kappa(\nabla\phi)^2$
for steep changes of composition.
 
For nonvanishing $\kappa$, the free energy functional \ref{esix}
supports static interfaces between phases which have small but finite
width, and gives an equilibrium interfacial tension that depends on
$\kappa$ and $F_0(\phi)$ (Bray 1994, Onuki 2002). Although interfacial
tension can thus be included without much modification into the
equations of motion, this would introduce a fourth dimensionless
control parameter into our model. We assume here that it plays no
major role and omit the corresponding terms from the equations of
motion, although in our preferred simulation algorithm for this type
of problem (lattice Boltzmann; see Kendon 2001), they are actually
incorporated in precisely the above way.

Evaluating \ref{diffusion} with $\kappa=0$ and 
$J_\phi = - M \nabla(\delta F/\delta\phi)$,  
introducing the abbreviations 
$\bar{D} \equiv M\, \left[  \delta^2 F_0[\phi]/\delta\phi^2  \right]_{\phi=\phi_0}$, 
$f(\phi) \equiv (M/\bar{D}) \; \delta^2 F_0[\phi]/\delta\phi^2$,  
and taking into account the temperature dependence of the factor
$\phi_0$ appearing in the dimensionless supersaturation $\hat\phi$ one
obtains equation \ref{etwo} for the evolution of $\hat\phi$.

\label{lastpage}
\end{document}